\documentstyle[prd,aps]{revtex}
\input psfig
\pssilent
\begin{document}
\title{The large cosmological constant approximation to classical and
quantum gravity: model examples}
\author{Rodolfo Gambini$^1$ and Jorge Pullin$^2$}
\address{1. Instituto de F\'{\i}sica, Facultad de Ciencias, 
Universidad
de la Rep\'ublica\\ Igu\'a esq. Mataojo, Montevideo, Uruguay}
\address{2. Center for Gravitational Physics and Geometry, Department
of Physics\\ The Pennsylvania State University, 104 Davey Lab,
University Park, PA 16802}
\date{Draft, Aug 14th  2000}
\maketitle
\begin{abstract}
We have recently introduced an approach for studying perturbatively
classical and quantum canonical general relativity. The perturbative
technique appears to preserve many of the attractive features of the
non-perturbative quantization approach based on Ashtekar's new
variables and spin networks.  With this approach one can find
perturbatively classical observables (quantities that have vanishing
Poisson brackets with the constraints) and quantum states (states that
are annihilated by the quantum constraints).  The relative ease with
which the technique appears to deal with these traditionally hard
problems opens several questions about how relevant the results
produced can possibly be. Among the questions is the issue of how
useful are results for large values of the cosmological constant and
how the approach can deal with several pathologies that are expected
to be present in the canonical approach to quantum gravity. With the
aim of clarifying these points, and to make our construction as
explicit as possible, we study its application in several simple
models. We consider Bianchi cosmologies, the asymmetric top, the
coupled harmonic oscillators with constant energy density and a simple
quantum mechanical system with two Hamiltonian constraints. We find
that the technique satisfactorily deals with the pathologies of these
models and offers promise for finding (at least some) results even for
small values of the cosmological constant. Finally, we briefly sketch
how the method would operate in the full four dimensional quantum
general relativity case.
\end{abstract}

\section{Introduction}
\subsection{The problem}

In the last few years, important developments in canonical quantum
gravity have taken place, culminating with the recent formulation of
two mathematically consistent \cite{Th,Vasi}, non-trivial canonical
quantizations of general relativity in terms of Ashtekar new variables
and spin networks. Instrumental in these developments have been the
underlying advances in mathematical techniques for dealing with
infinite dimensional nonlinear spaces, like the theory of cylindrical
functions and associated measures \cite{AsLe}, and the introduction of
spin networks to eliminate the over-completeness of the Wilson loop
basis \cite{RoSmvo} (for a recent summary, see the review article by
Rovelli \cite{Roliving}).  In spite of the advances, most of the
results obtained from these theories up to now concern statements made
{\em at a kinematical level} (without imposing the Hamiltonian
constraint). 

In order to start discussing if these theories contain the correct
semiclassical physics and are therefore physically viable theories of
quantum gravity, we need methods to introduce semiclassical states
(encouraging recent kinematical results can be seen in \cite{Thsemi})
but also to probe their dynamics.  In particular a current open issue
is if the semiclassical physics can be discussed at a kinematical
level only or requires the imposition of the correct dynamics.
Probing the dynamics has always been a problem in canonical general
relativity \cite{Thaw}. In any gauge theory, the only physically
relevant quantities are those that are invariant under the gauge
symmetries of the theory, or in the canonical language, that have
vanishing Poisson brackets with the constraints that represent the
gauge symmetries. At a quantum mechanical level, the physical states
are those that are annihilated by the constraints. In general
relativity we do not have any example of a quantity that classically
has vanishing Poisson brackets with the constraints \footnote{An
exception to this may be the quantity constructed in \cite{storna},
corresponding to the holonomy group of the spin connection.}, at least
in the case of a compact manifold. In fact, it is strongly suspected
that such quantities will never be constructed in closed form
\cite{torre}. Finding quantum states that are annihilated by the
constraints is also a challenge.

\subsection{An approximation scheme}

Most researchers in this area do not consider these problems a
fundamental obstruction, since in order to do meaningful physics one
only needs approximate expressions for the observables, as was
advocated long ago by Bergmann and Newman \cite{BeNe}. 
Finding an approximation method that does not destroy the
non-perturbative nature of the canonical treatment is, however, a
challenge.  We have recently proposed one such method \cite{GaPu00}.
It is based on considering general relativity coupled to a
cosmological constant and taking the $\Lambda\rightarrow\infty$
limit \cite{HuKu}. The construction goes as follows:

For general relativity with a cosmological constant, the Hamiltonian
constraint reads,
\begin{equation}
H(N)=H_{\Lambda=0}(N)+\Lambda \int d^3 x N(x) \sqrt{{\rm det} q(x)}
\end{equation}
where $N(x)$ is a smearing function, $H_{\Lambda=0}(N)$ is the
Hamiltonian constraint without a cosmological constant and ${\rm det}
q$ is the determinant of the spatial metric. If one now considers the
limit $\Lambda\rightarrow\infty$, and re-scales the constraint by
$1/\Lambda$, one is left with a theory, which we will call ``zeroth
order'' theory for which the Hamiltonian constraint is,
\begin{equation}
H^{(0)}(N)=\int d^3x N(x) \sqrt{{\rm det} q(x)},
\end{equation}
that is, the Hamiltonian constraint is just the square root of the
determinant of the spatial metric. In addition to this the theory has
the ordinary diffeomorphism constraint and if one uses Ashtekar
variables there will also be a Gauss law constraint (both are
independent of $\Lambda$) . Imposing classically the Hamiltonian
constraint of the zeroth order theory, one immediately sees that it
corresponds to metrics of identically vanishing determinant, that is,
degenerate metrics.

It is easy to construct quantities that have vanishing Poisson bracket
with the Hamiltonian constraint of the zeroth order theory. For
instance, one can consider any function depending only on the three
metric (or if one is using Ashtekar variables $(\tilde{E}^a_i,
A_b^j)$, the densitized triads $\tilde{E}^a_i$) and not on its
canonically conjugate momenta. To have a genuine observable the
quantity should also have vanishing Poisson bracket with the
diffeomorphism constraint. There are standard ways of achieving this
coupling the theory to matter \cite{obs,Smobs}. A novel point is that
in our approach, since the extra matter couplings in the Hamiltonian
are higher order in $\Lambda^{-1}$ these techniques yield genuine
observables for the zeroth order theory, whereas normally they just
construct quantities that have vanishing Poisson bracket only with the
diffeormophism constraint but not with the Hamiltonian
constraint. Since we will not need this technique for the examples we
consider in this paper, we do not include further details here 
(see a discussion in \cite{GaPu00}).

If one assumes that the observables of the theory are power series in
the inverse cosmological constant,
\begin{equation}
O_\Lambda(\pi,\tilde{E})=O^{(0)}(\pi,\tilde{E})+\Lambda^{-1}
O^{(1)}(\pi,\tilde{E})+\ldots,
\end{equation}
and one requests that these observables have vanishing
Poisson brackets with the Hamiltonian of the theory (more precisely,
one would like that the Poisson bracket be proportional to the 
constraints of the theory, for simplicity we will just demand that 
they vanish, but it is immediate to extend the construction to the
case in which they vanish on-shell)and we expand such
requirement in powers of $\Lambda^{-1}$ one gets
\begin{eqnarray}
\left\{O^{(0)},H^{(0)}\right\}&=&0\\
\left\{O^{(1)},H^{(0)}\right\}+\left\{O^{(0)},H^{(1)}\right\}&=&0.
\label{seconcon}
\end{eqnarray}

The first equation determines $O^{(0)}$ and the second one leads to a
linear partial differential equation for $O^{(1)}$. The construction
can be readily continued to higher orders. In all cases one obtains a
linear partial differential equation, albeit with a more and more
complex inhomogeneous term. It should be noted that one can obtain
many observables starting with different solutions to the first
equation. The linear partial differential equations are not hard to
solve, given that the coefficients of the derivatives are functions of
the triads, whereas the derivatives are with respect to the
connections. Given the simplicity of it, the system is always
integrable (although the solutions might be pathological, as we see in
the examples later in the paper) and therefore yields all the
observables of the theory.

This amazing simplicity also has a quantum counterpart, at the time of
finding states that are annihilated by the Hamiltonian constraint. If
we now focus on the formulation of the quantum constraints in terms of
the spin network representation of quantum gravity, for instance as
discussed by Thiemann \cite{Th}, one starts by considering our zeroth
order Hamiltonian, which is closely related to the well-understood
volume operator. It is easy to find eigenstates for this operator, in
particular ones with vanishing eigenvalues. When one considers the
first order corrections, because the zeroth and first order
Hamiltonians have a well defined algebraic action in the space of spin
network states, one gets a completely manageable (albeit complicated)
system of equations. We will not go into details of the quantum theory
of full general relativity in this paper, since it would make the
paper much less readable as we would have to introduce quite a bit of
(readily available but quite detailed) technology of spin networks.
We will only present, for completeness, a brief of what the issues in
the full quantum theory are in the last section of the paper.

Let us however describe in general lines how the quantum perturbative
approach would be for a generic system with a {\em single} Hamiltonian
constraint (we will discuss the case of more than one Hamiltonian as a
model of what can happen in a field theory in a subsequent section).

Consider a system where the quantum Hamiltonian constraint is of
the form $\hat{H}(\lambda)= \hat{H}^{(0)}+\lambda \hat{H}^{(1)}$ where
$\lambda$ is a parameter that we consider small. We then consider
quantum states that admit a power series expansion in powers of
$\lambda$,
\begin{equation}
|\psi(\lambda)> = |\psi^{(0)}>+\lambda |\psi^{(1)}> +\lambda^2
 |\psi^{(2)}>+\ldots 
\end{equation}

What one wants to do now is to solve the theory perturbatively. This
requires some comments. What does it mean to solve perturbatively a
quantum mechanical constrained system? The approach we will take is
the following: consider the eigenvalue equation for the Hamiltonian
constraint,
\begin{equation}
\hat{H}(\lambda) |\psi(\lambda)>=E(\lambda) |\psi>(\lambda),
\end{equation}
and consider its solution up to a certain order in $\lambda$, finding
the zero-energy eigenstate within such a theory.  That is, what we end
up doing is bound-state perturbation theory (provided the spectrum of
$\hat{H}^{(0)}$ is discrete) and finding one set of particular states,
those with zero eigenvalue. A good example to bear in mind about the
method is to consider an atom in a (weak) magnetic field and think
that one is looking for states with a given energy. If one goes to a
certain order in perturbation theory in terms of the magnetic field,
one will be able to achieve the given energy level only for certain
values of the magnetic field. That is, in this approach the
perturbative parameter is quantized. It is also clear that the
perturbed states that achieve a certain energy level for a given value
of the perturbative parameter do not necessarily stem from perturbing
the zeroth order theory level of that particular energy. It is
important to notice this in the gravitational case, where one looks
for states with vanishing eigenvalue. The perturbed states of
interest will in general not have $\hat{H}^{(0)}|\psi^{(0)}>=0$.

So we have cast the problem of solving quantum mechanically a
constrained system as finding perturbatively an eigenstate for a
given Hamiltonian in a quantum mechanical system. This problem is well
defined and, as we will see in the examples, solvable. For the full
general relativity case, the zeroth order Hamiltonian is quite
close to the volume operator, so we do have a discrete spectrum to
start with. However, because it is a field theory, one really has
infinitely many Hamiltonians and certain consistency requirements have
to be met. Therefore the quantum mechanical analogy we made may turn
out to be too na\"{\i}ve. The last example we consider in this paper
is geared towards clarifying that point.

Why should one care about a regime in which $\Lambda$ is large in
Planck units, given that the current observables value in such units
is $10^{-132}$?. Our attitude is the following: we have now at least
two candidates for theories of quantum gravity. We need to probe them
in any regime we can get a handle on in order to help decide if the
theories are plausible or not from a physical stand point. If we can
do this, we can then see in detail what regime of $\Lambda$ we can
really access. Perhaps it is a bit better than what one initially
expects and the results can be of interest to early universe
cosmology. Or perhaps one is lucky enough to encounter a situation
similar to other power series expansions (like the strong coupling
limit in the lattice or the large $N$ expansion in QCD) where the
results are robust and useful way outside the na\"{\i}vely expected
regime. We will see some hints of this in one of the examples of this
paper.

\subsection{Potential difficulties of the method and layout of 
answers provided in this paper}

The above observations make the whole approach worthwhile exploring
further. However, the surprising way in which it appears to deal
in a simple way with questions that in the full theory seem
unapproachable raises a certain level of skepticism. Several pointed
questions can be asked that make the whole approach look
questionable. Among them,

a) Does the approximation have any hope of working for values of
$\Lambda$ smaller than the Planck scale?

b) General relativity is a complex enough dynamical system, that some
pathologies might be expected. For instance, solutions might exhibit
some level of chaos. It could also happen that other pathologies
appear, like the full theory having less observables than the zeroth
order one. Does our method always produce $2 n-2m$ observables (where
$n$ is the number of degrees of freedom and $m$ the number of
constraints)? With a method that reduces everything to linear ODE's,
is there not a risk of ``papering over'' these subtleties.

c) What sort of approximation is one getting at a quantum level? What
is meant by an approximate state, or an approximate solution to a
quantum constraint?

d) When one applies the quantum perturbative approach in a field
theory, one effectively is dealing with infinitely many Hamiltonian
constraints. Is the method still viable or one gets an inconsistent
set of conditions on the perturbed states, leading to an empty theory?

e) The zeroth order Hamiltonian constraints we consider seem to admit
many more quantum states than the full Hamiltonian constraints. Does
the method produce spurious states for the full theory or correctly
notices the further limitations that should appear at higher orders?

This paper attempts to provide (at least partial) answers to these
questions by probing the perturbative method in model systems. In
section II we will consider the application of the technique to
Bianchi models. This will allow us to probe questions a) and b).
From Bianchi models we will learn that, surprisingly, when one
evaluates the approximate power series that we get for the observables
on solutions to the equations of motion, the cosmological constant
drops off from the expressions. Evaluated on the solutions to the
equations of motion, the expression for the observables become a sum
of terms that approximate better and better a constant of motion
without reference to the cosmological constant. For a Bianchi I 
recollapsing universe we will explicitly demonstrate that with only a
few terms one finds an expression that maintains an almost constant
(less than $10\%$ variation) value across $80\%$ of the life of the
universe. We will also see that in the case of Bianchi IX one can
easily construct observables, but it is likely that they will remain
good approximations only for short periods of time when one enters the
chaotic domain. We will also see that one does not obtain more
observables than the required ones. One can obtain many {\em
apparently different} power series with the method but they just
represent reparametrizations of the basic observables. We will also
learn how to deal with possible singularities in the expressions
obtained for the observables as functions of the phase space.

In section III we will discuss the heavy asymmetric top in three
dimensions. This is a Hamiltonian system that can be viewed as a
pathological constrained system if one considers it for a fixed value
of the energy. It is pathological in that it has less observables than
four.  We will see that our method {\em only apparently} generates
too many observables. On close inspection we will find that
several of the expressions are really not good candidates for
observables because they are expressions that have support only on
limited regions of phase space, that are in general not preserved by
evolution. In the end, the method produces the correct number of
observables for the model.

In section IV we explore a model consisting of two harmonic
oscillators with constant energy difference. This system has received
quite a bit of attention, for instance see \cite{oscil}. Unless the
ratio of the frequencies of the two oscillators is a rational number,
the system is pathological in that it only admits one quantum
state. We will show that our perturbative approach indeed notices this
fact and produces correct approximations for the quantum states in all
cases.

In section V we explore a model of coupled harmonic oscillators with
two ``Hamiltonian'' constraints. General relativity being a field
theory one has infinitely many Hamiltonian constraints (one per
spatial point). Therefore the other models we consider in the paper
seem a bit short in mimicking this feature, since they are mechanical
systems with only one constraint. In this section we increase modestly
this feature by considering a model with two Hamiltonians. We will see
that the method deals with no problem with such a system.

We will end with a short reflection on the implications of the results
of this paper and what challenges remain for the application of the
approximate scheme in the context of full classical and quantum
general relativity.

\section{Bianchi models}

As a first example of the application of the technique, we will
discuss the issue of finding observables for Bianchi models.  Bianchi
models are homogeneous cosmologies, and as a consequence the Einstein
equations become ordinary differential equations. This allows to
achieve, in some cases, quite a bit of progress towards their
solution. 

Following \cite{AsPu}, we represent the Bianchi models using as
variables the (diagonal) components of the triads $E^i$ and the
canonically conjugate momenta $A_i$. All class-A Bianchi models admit
a simple Hamiltonian formulation in terms of these variables, but we
will concentrate our attention in two models, the Bianchi I model and
the Bianchi IX model. The former is the simplest model and the latter
the one with the richest dynamics. Bianchi I models start from an
initial singularity and expand anisotropically.  Depending on the sign
of the cosmological constant they may expand forever or
recollapse. Bianchi IX models approach the singularity through an
infinite sequence of oscillations of increasing complexity.

In terms of the Ashtekar variables the Bianchi I and IX models 
with a cosmological constant are
described by a Hamiltonian constraint,
\begin{equation}
H= E^1 E^2 E^3 -{1 \over  \Lambda} \left(
E^1 E^2 (A_1 A_2 -\epsilon A_3)+
E^1 E^3 (A_1 A_3 -\epsilon A_2)+ 
E^2 E^3 (A_2 A_3 -\epsilon A_1) \right)
\end{equation}
where $\epsilon=1$ corresponds to the Bianchi IX model and
$\epsilon=0$ to the Bianchi I model. The Gauss law and diffeomorphism
constraints are identically satisfied.

To apply our construction we need to start by choosing an observable
for the zeroth order theory. In general such quantities are given by
$O^{(0)}=F(E^1,E^2,E^3,E^1 A_1-E^2 A_2,E^1 A_1-E^3 A_3)$. If one sets
out to find four independent observables one could start, for
instance, with the following independent choices for zeroth order
observables,
\begin{eqnarray}
O^{(0)}_1 &=& E^1\\ 
O^{(0)}_2 &=& E^2\\ 
O^{(0)}_3 &=& E^1 A_1-E^2 A_2\\ 
O^{(0)}_4 &=& E^1 A_1-E^3 A_3.
\end{eqnarray}
Let us now concentrate on the study of the Bianchi I cosmology. 
In this case $O^{(0)}_{3,4}$ are both exact observables for the
full model with a cosmological constant. Starting with $O^{(0)}_1$ one
finds the following first order correction,
\begin{equation}
O^{(1)}_1 = -{A_1 \over E^2 E^3}
\left(-E^1 E^2 A_2+(E^1)^2 A_1-E^1 E^3 A_3 \right),
\end{equation}
and in addition to this one has the solution to the homogeneous
equation, which is given by an arbitrary function that commutes with
$H^{(0)}$, that is, similar to $O^{(0)}$. The expression for
the Bianchi IX model is very similar to the one we just introduced,
it has an extra term $-{\epsilon A_1}$. 

In order to investigate the behavior of the approximate observable we
are considering we will evaluate it on an exact solution of the
equations of motion, the metric of reference \cite{swe},
\begin{equation}
ds^2 = -dt^2 +V(t)^{2/3} \left[
\Sigma(t)^{{4\over 3}\cos\left(\gamma\right)} dx^2+
\Sigma(t)^{{4\over 3}\cos\left(\gamma-{2 \over 3}\pi\right)} dy^2+
\Sigma(t)^{{4\over 3}\cos\left(\gamma+{2 \over 3}\pi\right)} dz^2
\right]
\end{equation}
where,
\begin{eqnarray}
V(t) &=& {\sin \omega t\over \omega}\\
\Sigma(t) &=&{2 \over \omega} {(1 -\cos\omega t)\over \sin \omega t}
\end{eqnarray}
and where $\omega=\sqrt{-3\Lambda}$ and where $\Lambda<0$. This
solution corresponds to a Bianchi I universe that expands out of a Big
Bang at $t=0$ but after a while the cosmological constant leads it to
recollapse at a time $\omega t_F = \pi$. The Ashtekar variables for
this metric (particularized to $\gamma=0$) read,
\begin{eqnarray}
E^1&=&\left (\cos(1/2\,\omega\,t)\right )^{4/3}\\
E^2&=&E^3={\frac {\sqrt [3]{2}\sqrt [3]{\tan(1/2\,\omega\,t)}
\left (\sin(\omega\,t)\right )^{2/3}}{\omega}}\\
A_1&=& 1/6\,{\frac {\sqrt [3]{2}\left (2\,\sin(\omega\,t)
\left (\cos(1/2\,
\omega\,t)\right )^{2}-3\,\sin(\omega\,t)-4\,\sin(1/2\,\omega\,t)\cos(
\omega\,t)\cos(1/2\,\omega\,t)\right )}{\sqrt [3]{\sin(\omega\,t)}
\left (\sin(1/2\,\omega\,t)\right )^{2/3}\left (\cos(1/2\,\omega\,t)
\right )^{2}}}\\
A_2&=& A_3=1/3\,{\frac {\sin(1/2\,\omega\,t)\omega}
{\sqrt [3]{\cos(1/2\,\omega\,t)}}}.
\end{eqnarray}

As we can see, for the solution in question $E_2$ and $E_3$ vanish at
$t=0$. If we look at the expression for the first order correction we
found, this implies that it diverges for $t=0$.  This is an example of
what we stated earlier, namely that the expressions for the
observables we found can sometimes be singular in certain points of
phase space. One can construct a first order correction that is well
behaved at $t=0$ by making use of the free function $F^{(1)}$ that
solves the homogeneous equation. If one adds to the first order
correction the solution to the homogeneous equation given by $(E^1
A_1-E^2 A_2)(E^1 A_1-E^3 A_3)(E^2 E^3)^{-1}$, it becomes,
\begin{equation}
O^{(1)}=A_2 A_3
\end{equation}
and this quantity is obviously well behaved at $t=0$. 

If one now proceeds to compute the second order correction, one finds
a similar situation with respect to the divergence at $t=0$. This can
be corrected by adding the function $F_2=
-1/3 \left((E^1 A_1-E^2 A_2) (E^1 A_1-E^3 A_3)\right)^2
(E^2 E^3)^{-2} (E^1)^{-1}$ and one obtains for the correction,
\begin{eqnarray}
O^{(2)}&=&
{1 \over 6 (E^2 E^3)^2 E^1}\left(
-2 (E^1)^2 A_1^2 E^2 E^3 A_2 A_3
+(E^1)^2 A_1^2 A_3^2 (E^3)^2\right.\\
&&-2 E^1 A_1 E^2 (E^3)^2 A_2 (A_3)^2
+(E^1)^2 A_1^2 A_2^2 (E^2)^2\\
&&\left. 
-2 E^1 A_1(E^2)^2 E^3 A_2^2 A_3-2 A_2^2 (E^2)^2 A_3^2 E_3^2\right)
\nonumber
\end{eqnarray}
and a similar subtraction yields the third order correction,
\begin{eqnarray}
O^{(3)}&=&
-{1\over 30 (E^2 E^3)^3(E^1)^2}
\left(-6 (E^1)^3 A_1^3 (E^2)^2 E^3 A_2^2 A_3
-6 (E^1)^3 A_1^3 E^2 (E^3)^2 A_2 A_3^2
-12 E^1 A_1 (E^2)^3 (E^3)^2 A_2^3 A_3^2\right.\\
&&-12 E^1 A_1 (E^2)^2 (E^3)^3 A_2^2 A_3^3
-14 (E^1)^2 A_1^2 (E^2)^2 (E^3)^2 A_2^2 A_3^2
-3 (E^1)^2 A_1^2 E^2 (E^3)^3 A_2 A_3^3\nonumber\\
&&-3 (E^1)^2 A_1^2 (E^2)^3 E^3 A_2^3 A_3
-4 (E^1)^4 A_1^4 A_3 E^3 A_2 E^2
+2 (E^1)^4 A_1^4 A_3^2(E^3)^2
+2 (E^1)^4 A_1^4 (E^2)^2 A_2^2\nonumber\\
&&-6 A_2^3 (E^2)^3 A_3^3 (E^3)^3
+6 A_2^3 (E^2)^3 (E^1)^3 A_1^3
+6 (E^1)^3 A_1^3 A_3^3 (E^3)^3)\nonumber
\end{eqnarray}
If we now evaluate the above expressions for the exact solution we 
discussed above one gets ($x=\omega t$),
\begin{eqnarray}
O^{(0)}_1 &=&\cos^{4/3}({x\over 2})\\
{O^{(1)}_1 \over \Lambda} &=&
1/3\,{\frac {2\cos^{2}({x\over 2})+1}
{\cos^{2/3}({x\over 2})
}}\\
{O^{(2)}_1 \over \Lambda^2} &=&
1/27\left[
-19\cos^{4}({x\over 2})
+19\cos^{4}({x\over 2})\cos^{2}(x)
-6\cos^{2}({x\over 2})
+6\cos^{2}({x\over 2})\cos^{2}(x)\right.\\
&&\left.
-4\,\cos({x\over 2})\sin(x)\cos(x)\sin({x\over 2})-2
+4 \cos^{3}({x\over 2})\sin(x)\cos(x)\sin({x\over 2})
+2\cos^{2}(x)\right]
\left[\cos^{8/3}({x\over 2})
\left(-1+\cos^{2}(x)\right )\right]^{-1}\nonumber\\
{O^{(3)}_1 \over \Lambda^3} &=&
{\frac {1}{810}}
\left[
-568\cos^{6}({x\over 2})-15
+15\cos^{2}(x)
-194\cos^{4}({x\over 2})
+488\cos^{6}({x\over 2})\cos^{2}(x)
+354\cos^{4}({x\over 2})\cos^2(x)\right.\nonumber\\
&&\left.-47\cos^{2}({x\over 2})\cos^2(x)
-33\cos^2({x\over 2})
-16\cos^3({x\over 2})\sin(x)\cos(x)
\sin({x\over 2})
-72\cos({x\over 2})\sin(x)\cos(x)\sin({x\over 2})\right.\nonumber\\
&&\left.+88\cos^5({x\over 2})\sin(x)\cos(x)
\sin({x\over 2})\right]
\left[\cos^{14/3}({x\over 2})
\left(-1+\cos^{2}(x)\right )\right]^{-1}.
\end{eqnarray}

The first observation is that the above expressions for the correction
have lost the dependence on $\Lambda$ stemming from the perturbative
approach (there is a trivially re-scalable dependence on $\Lambda$
through $x=\omega t$). Therefore, if the above expressions converge,
the convergence will be independent of $\Lambda$. This can be seen in
the following plot, in which we show the observables as functions of
$x$. We see that in spite of the lack of dependence in $\Lambda$, the
curves do converge, the third order one maintaining an almost constant
value for the longest period of time (more than half the lifetime of
the universe). The various corrections are not well behaved near the
big crunch.

\begin{figure}
\centerline{\psfig{figure=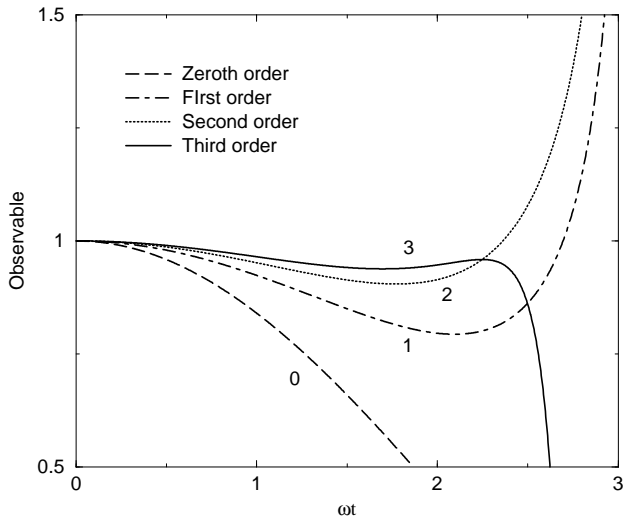,height=70mm}}
\caption{The perturbative observable evaluated for a Bianchi I
cosmology with negative cosmological constant. The time runs from
zero, the Big Bang, to $\pi$, the Big Crunch.  The four curves
correspond to better and better approximations.}
\end{figure}

It should be remembered that the above calculations were
particularized for $\gamma=0$. The value of the observable will in
general be $\gamma$-dependent. One way of understanding this is that
the ``physical'' information content of the observable is to give a
measure of the $\gamma$ parameter, which is associated to the
anisotropy of the model. For instance, the exact observable $O_3=A^1
E^1-A^2 E^2$ takes the value $O_3=-\cos\gamma +{\sqrt{3} \over
3}\sin\gamma$. Since the solution to the equations of motion we are
considering is parameterized by a single constant, all observables
become given functions of that constant.

As was shown in the above calculations, the Bianchi IX model 
appears as only slightly more complex, and similarly any class-A
Bianchi model can be treated in the same way. Because of the local
nature of our calculations (in time), it is not surprising to find
observables in the Bianchi IX case. In that cosmology, because of the
chaotic oscillations the dynamics undergoes, it is unlikely one will
be able to find expressions for the observables that are global, but
expressions like the ones we find should face no difficulty.

As we saw, the zeroth order observable one starts from has a huge
degree of ambiguity. There is also a lot of ambiguity in each
perturbative step, due to the presence of the solution of the
homogeneous equation. It is evident that certain choices will lead to
more useful (or in some cases shorter or even polynomial) expressions,
as we exhibited when we manipulated the homogeneous solutions to avoid
expressions that are singular at certain points in phase space.

Summarizing, we learn the following lessons from the Bianchi example
we have just discussed. First of all, we see that the approximation
for the observables we get, when evaluated on a solution of the
equations of motion, becomes independent of the cosmological
constant. This suggests that the perturbative approach will remain
good even when the constant is small. In fact, we see in the example,
that the observable is well approximated for a significant fraction of
the lifetime of the universe in question. We also see that the
presence of chaos in the Bianchi IX model does not preclude us from
finding approximate expressions for the observable, though it is
likely that it will impose further restrictions on the domain of
validity (in terms of evolution time) of the approximate observable. 
It is possible to test these hypotheses with numerical evolutions of
the Bianchi IX cosmology.

\section{The asymmetric heavy top with a fixed energy value}

As another example of the application of the perturbative Hamiltonian
ideas we are proposing, we analyze a model which in spite of its
simplicity, adds pathological behaviors that are interesting to
analyze under the proposed scheme. The model is an asymmetric heavy
top with constant energy. This system only has one observable, the
momentum canonically conjugate to the angle around the vertical axis.
We will assume that the asymmetry in the top is small and treat this
model as a perturbation of a symmetric heavy top, which has two
observables. Will our technique notice that one loses an observable
when one loses the symmetry?

Consider a heavy symmetric top with moments of inertia $I_1=I_2$ and
$I_3$. Now suppose one sticks on two small masses at opposite sides of
the surface of the top, forming a line perpendicular to and
intersecting the top's third axis at its center of mass. Then the
moments of inertia along directions $1$ and $2$ will be slightly
different. Let us quantify this as $I_2=I_1+\lambda$, with $\lambda$ a
small perturbation (we use the lowercase to avoid confusion with the
cosmological constant, although the parameter plays the same
perturbative role in this problem as the inverse cosmological constant
in full general relativity).

The Lagrangian for such a system will be given, in terms of the
traditional Euler angles (see for instance \cite{Goldstein}), 
\begin{equation}
L={1 \over 2} I_1 \left(\dot{\theta}^2+\dot{\varphi}^2 \sin^2
\theta\right) +{1 \over 2} I_3 \left(\dot{\psi}^2+\dot{\varphi}^2
\cos\theta\right)^2 +{\lambda\over 2}\left(\dot{\varphi}\sin\theta\cos\psi-
\dot{\theta}\sin\psi\right)^2 -M g \ell \cos\theta
\end{equation}
where $\ell$ is the length along the top third axis 
from the top base up to the center of mass. To introduce a canonical
formulation, we compute the canonical momenta,
\begin{equation}
p_\theta\equiv {\partial L \over \partial \dot{\theta}}=
I_1 \dot{\theta}+\lambda\dot{\theta}\sin^2\psi-
\lambda \dot{\varphi}\sin\theta\cos\psi\sin\psi
\end{equation}
and similar equations for the other momenta, 
which yields an invertible linear system of equations from where one
can compute the velocities as functions of the canonical momenta.  One
can then perform a Legendre transform and obtain a Hamiltonian
$H(p_\theta,p_\varphi,p_\psi,\theta,\psi,\lambda)$. For small values
of the perturbative parameter, one can expand,
\begin{equation}
H=H(\lambda=0)+ \lambda {\partial H \over \partial \lambda}(\lambda=0)\equiv
H_0+\lambda H_1,
\end{equation}
where the general form of the Hamiltonians is,
\begin{eqnarray}
H_0&=&{(p_\theta)^2\over 2 I_1} + \sum_{i,j=2}^3 c_{ij}(\theta) p_i
p_j,\\ H_1&=&=\sum_{i,j=1}^3 d_{ij}(\theta,\psi) p_i p_j
\end{eqnarray}
with $p_1=p_\theta, p_2=p_\varphi, p_3=p_\psi$.
We would like to treat the system as an example of constrained
system. Therefore we request that the energy have a given value
$H=E_0$, that is,
\begin{equation}
H^{(0)}-E^{(0)}+\lambda H^{(1)}=0. \label{31}
\end{equation}

Before treating the case of the top with constant energy, let us first
recall how one formulates the usual asymmetric top, but as a
parameterized system. This allows a more natural contact with the
constrained systems we will handle later.  The parameterized top is
obtained by introduce a time $q^0$ and its canonically conjugate
momentum $p_0$, such that,
\begin{equation}
p_0-H(p_i,q^i)=0
\end{equation}
where $q^1=\theta, q^2=\varphi, q^3=\psi$.  
The observables are defined by,
\begin{equation}
\left\{ O,p_0-H(p_i,q^i)\right\}=0,
\end{equation}
which yields,
\begin{equation}
{\partial O \over \partial q^0}-{\partial O \over \partial q^i}
{\partial H \over \partial p_i}+{\partial O \over \partial p_i}
{\partial H \over \partial q^i}=0.
\end{equation}
This system is solved by finding the solution to the following set of
differential equations,
\begin{equation}
{d q^0 \over d s}=1, {d q^i \over ds}=-{\partial H \over \partial
p_i},
{d p_i \over ds}={\partial H \over \partial q^i}.
\end{equation}
The first of these equations can be integrated immediately, $q^0=s+C$.
The other two equations are just Hamilton's equations of motion, which
integrated yield,
\begin{equation}
q^i=f^i(p^0_i,q^i_0,s),  p_i=g_i(p^0_i,q^i_0,s), 
\end{equation}
if one substitutes $s$ by $-q^0$ in the above expressions, one obtains
the six constants of motion,
\begin{equation}
q^0_i=f_i(p_i,q^i,-q^0),\quad p_0^i=g_i(p_i,q^i,-q^0),
\end{equation}
which are observables of the parameterized system. A crucial element
in the solution of this system, that allows to obtain the six
constants of motion explicitly is that the solution of the equation
for $s$ in terms of $q^0$ is straightforward. We will see that this
poses problems in the next example.

Let us now study the case of interest, the (slightly) asymmetric heavy
top, with Hamiltonian constraint (\ref{31})
and assuming observables of the form $O=O_0+\lambda O_1+\ldots$. Let
us start with the equations to zeroth order in $\lambda$,
\begin{equation}
\left\{O^{(0)},H^{(0)}-E^{(0)}\right\}=0,
\end{equation}
with 
\begin{equation}
H^{(0)}={(p_1)^2\over 2 I_1}+\sum_{i,j=2}^3 c_{ij}(\theta)p_i p_j,
\end{equation}
which leads to the PDE,
\begin{equation}
{p_1 \over I_1} {\partial O_0 \over \partial q^1}+\sum_{i,j=2}^3
c_{ij}(\theta)p_j {\partial O_0\over \partial q^i}+ 
{\partial O_0\over \partial p_1}
\sum_{i,j=2}^3 {d c_{ij}\over d\theta} p_i p_j=0
\end{equation}
which again can be solved by the substitution technique,
\begin{eqnarray}
{d q^1 \over d s} &=& {p_1 \over I_1}, {d q^i\over ds} = \sum_{i,j=2}^3
c_{ij}(\theta) p_j,\quad i=2,3\\
{d p_1 \over d s} &=& \sum_{i,j=2}^3 c'_{ij} p_i p_j, {d p_i \over ds}=0,\quad i=2,3.
\end{eqnarray}

At zeroth order the system has two trivial constants of motion,
\begin{equation}
p_i=p^0_i, \quad i=2,3.
\end{equation}
One can in principle integrate the remaining equations to yield,
\begin{eqnarray}
\theta&=&\theta(p_i^0,q^i_0,s),p_\theta=p_\theta(p_i^0,q^i_0,s)\\
\varphi&=&\varphi(p_i^0,q^i_0,s),\psi=\psi(p_i^0,q^i_0,s),\quad
i=1,3.
\end{eqnarray}
As before, one can (formally) invert the solutions to obtain the
constants of motion as functions of phase space by solving for $s$ and
substituting back in the following equations:
\begin{eqnarray}
q^i_0&=&f_i(p_i,q^i,s), p_i^0=g_i(p_i,q^i,s),\quad i=1,3\\
p_i&=&p^0_i, \quad i=2,3.
\end{eqnarray}
We therefore appear to have found again six constants of motion as in
the case we studied before. However, the above equations involve
periodic functions with arguments with incommensurate periods. One
will only determine $s$ up to multiples of the period and when
substituting in the other equations will end up with multivalued
functions. If one chooses a particular branch, the resulting
observables only take values on a portion of the accessible phase
space. They are therefore not acceptable as observables for the 
system.

This is an important observation, since we will see that this is the
mechanism which prevents our technique from generating observables for
systems that do not have them.

Let us now consider the first order corrections. 
The equation to be solved are,
\begin{equation}
\left\{O_1,H_0-E_0\right\}+\left\{O_0,H_1\right\}=0
\end{equation}
and if we start from $O_0=p_\psi$, we are left with,
\begin{equation}
{p_1\over I_1}{\partial O_1\over\partial q^1}+\sum_{i,j=2}^3 
c_{ij}(\theta) p_j {\partial O_1\over \partial q^i}
+{\partial O_1\over\partial p_1} \sum_{i,j=2}^3 
c'_{ij}(\theta) p_j =
\sum_{i,j=2}^3 {\partial d_{ij}\over \partial \psi} p_i p_j.
\end{equation}
Now the PDE is non-homogeneous. We can still solve it with our
parameterization technique via,
\begin{eqnarray}
{d q^1 \over ds} &=&{p_1 \over I_1},\\
{d q^i\over ds}&=&\sum_{i,j=2}^3c_{ij} p_j,\quad i=2,3\\
{d p_1\over ds}&=&\sum_{i,j=2}^3c'_{ij} p_i p_j,\quad i=2,3\\
{d p_i\over ds}&=&0\\
{d Z\over ds}&=&\sum_{i,j=2}^3 {\partial d_{ij}(\theta,\psi)\over
d\psi} p_i p_j\quad i=1,3.
\end{eqnarray}
As before, we will need to solve for $s$ and replace it in the
equations to obtain expressions for the initial values as functions of
phase-space. We do run, however, into the same difficulty as before,
we cannot obtain a uniquely defined expression for $s$ and therefore
we cannot construct a first order correction for the observable
chosen. The procedure therefore fails to work.

What happens if we choose $O_0=p_\varphi$ instead? In that case
$\{O_0,H_1\}=0$ and therefore the equation for $O_1$ indeed has a
solution. We will therefore recover, order by order, an expression for
the {\em single} observable the system admits.

The general conclusion seems to be that, although the linear PDE's of
our procedure in principle can always be solved, requiring that the
solution be a well defined function on phase space (and not only on a
portion of it) severely restricts the observables that can be obtained
by the technique, and apparently prevents the technique from
generating spurious observables in systems with less observables than
degrees of freedom.

\section{Two oscillators with constant energy difference}
\subsection{The model}
Let us now consider examples of the application of the method at a
quantum level. Haj\'{\i}\v{c}ek and others \cite{oscil} have
considered a model of constrained system, consisting of two harmonic
oscillators with a fixed energy difference. The possible eigenstates
of the system therefore consist (if one uses the standard inner
product for the oscillators) in situations in which oscillators have
the appropriate quanta of energy to satisfy the constraint that the
difference of their energies be constant. It is clear that this can
only happen if the ratio of the frequencies of the oscillators is
rational. Otherwise, the system would admit at most one quantum
state. This is the pathology of this system: if the ratio of the
frequencies is not rational, the quantum system only has one state
satisfying the energy constraint. If it is rational, the system admits
infinitely many states. One could therefore consider starting with a
zeroth order model consisting of two coupled oscillators with rational
ratio of frequencies and perturb the system by adding the Hamiltonian
of two oscillators with an irrational ratio. The zeroth order system
has infinitely many quantum states and the perturbed system has only
one state. Can our method detect and correctly handle this situation?
This is what we attempt to probe in this section.

\subsection{Zeroth order Hamiltonian}

The Hamiltonian constraint for the system in question simply
reads,
\begin{equation}
H={1 \over 2} (p_1^2+\omega_1^2 x_1^2)
-{1 \over 2} (p_2^2+\omega_1^2 x_2^2)+1=0
\end{equation}
where we have chosen the energy difference equal to one. The
quantization of this model, viewed as a constrained system, has order
ambiguities. Consider the transformation
\begin{equation}
\bar{x}_i=\sqrt{\omega_i\over \hbar} x_i,\quad
\bar{p}_i=\sqrt{\hbar\over \omega_i} p_i,
\end{equation}
the constraint reads,
\begin{equation}
H={1 \over 2}\hbar \omega_1 (\bar{p}_1^2+\bar{x}_1^2)
-{1 \over 2}\hbar \omega_1 (\bar{p}_2^2+\bar{x}_2^2)+1 =0
\end{equation}
and introducing the complex variables,
\begin{equation}
\alpha_j={1 \over \sqrt{2}} (\bar{x}_j+i\bar{p}_j),\quad
\alpha_j^*={1 \over \sqrt{2}} (\bar{x}_j-i\bar{p}_j),
\end{equation}
the constraint takes the form,
\begin{equation}
H=\hbar \omega_1 \alpha_1^* \alpha_1 -
\hbar \omega_2 \alpha_2^* \alpha_2 +1=0.
\end{equation}

We can then proceed to the quantization,
\begin{eqnarray}
\bar{x}_j\rightarrow\hat{\bar{x}}_j,&\quad&
\bar{p}_j\rightarrow\hat{\bar{p}}_j\\
\left[\hat{\bar{x}}_j,\hat{\bar{p}}_k\right]{}&=&i \delta_{jk}\\
\alpha_j\rightarrow \hat{a}_j,&\quad&
\alpha^*_j\rightarrow \hat{a}^\dagger_j,
\end{eqnarray}
and the quantum constraint can be written as
$\hat{H}=\hat{H}_1-\hat{H}_2+1=0$,
with,
\begin{equation}
\hat{H}_j=\hbar \omega_j \left(\mu_j \hat{a}^\dagger_j \hat{a}_j +
(1-\mu_j) \hat{a}_j\hat{a}^\dagger_j\right)
\end{equation}
with $\mu_j$ representing the ordering ambiguities. To obtain the
quantum states that are annihilated by this constraint, let us
consider separately the spectra of both $\hat{H}_j$'s. Since they are
harmonic oscillators, we have,
\begin{equation}
H_j=\hbar\omega_j n_j +\hbar K_j, \qquad {\rm eigenvector}\quad |n_j>
\end{equation}
with $K_{1,2}$ arbitrary ordering-dependent constants. In order to
solve the constraint, the values of $n_1,n_2$ must satisfy,
\begin{equation}
\hbar \omega_1 n_1+\hbar K_1-\hbar \omega_2 n_2 -\hbar K_2 +1=0.
\end{equation}
To produce solutions, one can choose the constants $K_j$. For
instance, for $n_1=n_2=0$, we get $\hbar(K_1-K_2)+1=0$.

In order to have more than one solution, one needs to have,
\begin{equation}
\omega_1 n_1-\omega_2 n_2=0,
\end{equation}
or,
\begin{equation}
{n_1 \over n_2} ={\omega_2 \over \omega_1},
\end{equation}
that is, the ratio of the frequencies of the oscillators 
must be a rational number. If ${n_1 \over n_2}={N \over M}$ then
$n_1 = N r$, $n_2=M r$ with $r$ an integer and the wavefunctions of
the physical space are $|Nr,Mr>=|Nr>|Mr>$.

\subsection{Perturbed model: exact treatment}

We will now construct a model that is a perturbation of the above
considered one but is still exactly solvable. We can then solve the
model exactly and perturbatively and compare the results.
Let us consider the system \footnote{Notice that this is equivalent,
up to an overall rescaling, to add a perturbation corresponding to the
energy difference of two oscillators, as we proposed doing in the
introduction.},
\begin{equation}
H=H_1-H_2-\lambda {\omega_2^2\over 2} x_2^2+1=0.
\end{equation}

This system is equivalent to the original Hamiltonian constraint if we
define ${\omega'}_2=\omega_2 (1+\lambda)^{1/2}$. We can therefore
readily solve it. For 
system to have more than one state, we have,
\begin{equation}
\omega_1 n_1 +\omega'_2n_2=0,
\end{equation}
or, 
\begin{equation}
{n_1 \over n_2}={\omega'_2\over \omega_1}={\omega_2
\sqrt{1+\lambda}\over \omega_1},
\end{equation}
therefore $n_1=P r$ and $n_2 =Q r$ with $r$ integer, and 
remembering that ${\omega_2\over\omega_1}={N\over M}$,
in order to have a solution the perturbative parameter has to have a
given value,
\begin{equation}
\lambda={P^2 M^2 \over Q^2 N^2}-1.\label{1}
\end{equation}

In order to compare with the perturbative calculation we will perform
in the next paragraphs, let us consider the condition for having more
than one solution  for this model,
\begin{equation}
\omega_1 n_1 -\omega'_2 n_2=0=\omega_1 n_1 -\sqrt{1+\lambda}\omega_2 n_2
\end{equation}
which in the case $\lambda\ll 1$ reads, to first order,
\begin{equation}
\omega_1 n_1 -\left(1+{\lambda\over 2}\right)\omega_2 n_2=0,
\end{equation}
or,
\begin{equation}
\lambda=2\left({M\over N}{n_1\over n_2} -1\right).
\end{equation}
In order to have $\lambda \ll 1$, the ratio $n_1/n_2$ must be a
rational number close to $N/M$. We can therefore write it as,
\begin{equation}
{n_1 \over n_2} = {N r+ \ell \over M r + m},
\end{equation}
with large $r$. Expanding, we therefore get,
\begin{equation}
\lambda=2 \left[{r+ {\ell \over N}\over r+{m\over M}}-1\right]\sim
 2 {\ell \over N r} -2 {m\over M r} + O(r^{-2}).
\end{equation}

Summarizing, the perturbed model can have infinitely many or a single
quantum state, depending on the value of $\lambda$. Let us now see if
treating the model perturbatively we recover the same result.

\subsection{Perturbative treatment of the model}

We consider the zeroth and first order Hamiltonians as,
\begin{eqnarray}
H^{(0)}&=&H_1-H_2+1\\ 
H^{(1)}&=&-{1\over 2} \omega_2^2 x_2^2=-{1\over 2}
\hbar \omega_2^2 \bar{x}_2^2.
\end{eqnarray}

We start by considering the eigenvalue problem up to first
order in $\lambda$, 
\begin{equation}
\left(<\phi^{(0)}+\lambda \phi^{(1)}|\right)
\left(\hat{H}^{(0)}+\lambda
\hat{H}^{(1)}\right) = \left(\epsilon^{(0)}+\lambda \epsilon^{(1)}\right)
\left(<\phi_0+\lambda \phi_1|\right).
\end{equation}
Let us start by considering the resulting zeroth order equation,
\begin{equation}
<\phi^{(0)}|H^{(0)} = \epsilon^{(0)} <\phi^{(0)}|.
\end{equation}
The spectrum of $H^{(0)}$ is,
\begin{equation}
\epsilon_{n_1,n_2}^{(0)}=\hbar \left(\omega_1 n_1 
-\omega_2 n_2+K_1-K_2\right)+1.
\end{equation}
Here we need to distinguish if $\omega_1/\omega_2$ is rational or
not. If it is rational, then the spectrum is degenerate and we need to
apply ordinary degenerate bound state perturbation theory as described
in any quantum mechanics textbook. If it is not rational, the spectrum
is not degenerate. As we discussed in the introduction, let us 
concentrate on the case in which
the frequencies are related by $\omega_2/\omega_1=N/M$. 
The degeneracy of the spectrum can be directly seen in that 
it takes the same values for $n_1 = p +N r$, 
$n_2=q+M r$ with given $p,q$ and arbitrary values of $r$, that is, 
\begin{equation}
\epsilon^{(0)}_{p,q} = \hbar \left( \omega_1 p -\omega_2 q
+K_1-K_2\right)+1 
\end{equation}
The eigenstates of $\hat{H}^{(0)}$ are,
\begin{equation}
<\phi_{p,q,r}|=<p+N r,q+M r|
\end{equation}
in the number representation for the harmonic oscillators. We
therefore find that the solution to zeroth order is given by (at the
moment) an arbitrary combination of the given eigenstates,
\begin{equation}
<\phi^{(0)}_{p,q}| = \sum_r C(r) <\phi_{p,q,r}|.
\end{equation}

Let us now consider the equation to first order in $\lambda$,
\begin{equation}
<\phi^{(0)}_{p,q}| H^{(1)} + <\phi^{(1)}_{p,q}| H^{(0)}=
\epsilon^{(0)}_{p,q} <\phi^{(1)}_{p,q}|+\epsilon^{(1)} 
<\phi^{(0)}_{p,q}|.\label{firstlambda}
\end{equation}

Since $\left(<\phi_0|+\lambda <\phi_1|\right)\left(|\phi_0>+\lambda
|\phi_1>\right)=1$, to first order in $\lambda$ and we can choose the
phase of $|\phi_1>$ to have,
\begin{equation}
<\phi_0|\phi_0>=1,\quad <\phi_0|\phi_1>=<\phi_1|\phi_0>=0 \forall p,q.
\end{equation}
We now take the equation (\ref{firstlambda}) and project it onto the
the eigenspace with eigenvalue $\epsilon^{(0)}_{p,q}$. While doing
this, the contribution from $<\phi^{(1)}_{p,q}| H^{(0)}$ acting 
on this space cancels with the term $\epsilon^{(0)}_{p,q}
<\phi^{(1)}_{p,q}|$ acting on the same space. We are therefore left
with, 
\begin{equation}
<\phi^{(0)}_{p,q}|\hat{H}^{(1)}|p+Nr,q+M r> = 
\epsilon^{(1)} <\phi^{(0)}_{p,q}|p+Nr,q+M r>,
\end{equation}
and recall that $\hat{H}^{(1)}=1/2 \hbar \omega_2 
\hat{\bar{x}}_2^2$ and
in turn,
\begin{equation}
\hat{\bar{x}}_2^2 = {\left(\hat{a}_2+\hat{a}^\dagger_2\right)^2\over
2}= {\hat{a}_2^2 +\left(\hat{a}_2^\dagger\right)^2+ \hat{a}_2
\hat{a}_2^\dagger + \hat{a}_2^\dagger \hat{a}_2\over 2} = 
{\hat{a}_2^2 +\left(\hat{a}_2^\dagger\right)^2+ 2 \hat{a}_2^\dagger
\hat{a}_2+1 \over 2}.
\end{equation}

Since the matrix elements of $H^{(1)}$ are between states of energy
$\epsilon^{(0)}_{p,q}$, $\hat{a}_2^2$ and $(\hat{a}_2^\dagger)^2$ do
not contribute and the Hamiltonian is diagonal. Therefore of all the
terms in the superposition defining $<\phi^{(0)}|$ we are left with
$<p+Nr,q+Mr|$, which normalizing yields,
\begin{equation}
<\phi_0|=<p+Nr,q+M r|
\end{equation}
and energy,
\begin{equation}
\epsilon^{(1)} = -{1 \over 2}\hbar \omega_2 
<\phi_0| {2 \hat{a}_2^\dagger \hat{a}_2+ 1 \over 2}|\phi_0> = 
-{1 \over 2} \hbar \omega_2 \left( q + M r + {1 \over 2}\right).
\end{equation}
Therefore the first order corrected energy is,
\begin{equation}
\epsilon^{(0)}+\lambda \epsilon^{(1)} = \hbar \omega_2 \left( {M\over
N} p -q -{\lambda \over 2} \left(q + M r +{1 \over 2}\right)\right)
+\hbar\left(K_1 -K_2\right)+1.
\end{equation}
and as before we need to 
choose $K_1, K_2$ in such a way that $|0,0>$ is a solution
of the theory up to the order we are considering,
i.e., $K_1-K_2+1 -{\lambda \over 4} \hbar \omega_2 =0$. If we want
other solutions, we need to make the energy eigenvalue vanish, up to
the order we are considering this implies,
\begin{equation}
\lambda=2 {{M\over N}p -q\over q+M r }.
\end{equation}

For small values of $\lambda$ we should recover the results of the
exact calculation. We notice that $\lambda$ decreases with increasing
$r$. In that limit we therefore have that,
\begin{equation}
\lambda\sim {2 \over r} 
\left({p \over N}-{q \over M}\right) +O(r^{-2}),\label{valuelambda}
\end{equation}
and this agrees with the expansion in small $\lambda$ of our exact
calculation. 

It should be noted that the above calculation shows that starting from
a given state in the zeroth order theory and choosing a perturbative
parameter, we get a first order solution to the constraint. It is
immediate to see however, that this can be accomplished by starting
from an infinite number of sets in the zeroth order theory. Simply
consider a state obtained by multiplying $p,q$ and $r$ times an
integer. The same value of lambda in equation (\ref{valuelambda})
ensures that the first order state is a solution.

The calculation can be completed by considering the projection of
(\ref{firstlambda}) on the subspace of states orthogonal to the states
of energy $\epsilon^{(0)}_{p,q}$. Such projection determines the
correction to the state, $<\phi^{(1)}|$. This can be straightforwardly
done.

The calculation can be repeated for irrational quotients of
frequencies. In that case one is doing ordinary (non-degenerate) bound
state perturbation theory. The calculations resemble very much the
ones we did above, so we will not detail them here. Some comments are
nevertheless in order.

The zeroth order theory has only one state when the frequency ratio is
irrational and many states when it is rational. In the perturbed
theory, in both cases we can find values of $\lambda$ such that the
theory has many states. That is, the perturbed model we constructed is
such that depending on the value of the perturbative parameter one has
either one or many states. If we do not choose the correct value of
lambda (as determined by equation (\ref{valuelambda})), then the model
only admits the state $|0,0>$ as solution.

In particular, we have correctly addressed the situation posed in the
introduction to this section: if the zeroth order model has rational
frequencies (and therefore infinite solutions) but the peturbed model
corresponds to a value of $\lambda$ that only admits one solution, we
will not satisfy the energy constraint to first order and we are
therefore left with the vacuum as the only solution (in the factor
ordering chosen). Therefore the method seems to handle well the
drastic reduction in number of states implied by the perturbation.

It is suggestive that the model considered has a quantized value of
$\lambda$ in the full treatment if one wishes to have infinitely many
states. That behavior is reproduced correctly in the perturbative
approach. However, the quantization of the perturbative parameter is a
feature of the perturbative method as long as the zeroth order
Hamiltonian has a discrete spectrum (which is the only case in which
we can apply the method). It is interesting to notice that in the
other models considered by Hajicek, the spectrum of the Hamiltonian is
continuous and therefore we cannot apply our method, and the full
models do not have any particular requirements on the frequencies of
the oscillators.

Another lesson from this example is that the lack of states in the
perturbative treatment strongly suggests that the full model also
lacks quantum states, unless there is some sort of non-analytical
behavior in terms of the perturbative parameter.

\section{A model with two Hamiltonians}

In the case of quantum gravity, one is dealing with a field
theory, therefore one has an infinite number of constraints. In the
spin network representation, if one considers networks with a finite
number of vertices, the number of equations resulting from the
constraints might be finite, but in realistic situations of
semi-classical interest it will still be large. Therefore the quantum
models we have considered up to now, which are quantum mechanical
systems with one constraint, may not capture some of the aspects
present in the gravitational case. It might be that dealing with a
larger than one number of constraints implies further relations
between the perturbative coefficients that make the system
incompatible. It is somewhat unlikely that this will happen, since
after all, assuming the full theory is solvable, if the solution
admits a power series expansion in $\Lambda$, one should find a
solution that is acceptable perturbatively to any order in
perturbation theory. Nevertheless, it might be useful to consider a
simplified model with more than one constraint to make sure the
technique works. It is in particular interesting to see how, if one
has several constraints to satisfy, one can achieve it with the same
value of the perturbative parameter for all the equations. 
This is what we attempt in this section.

\subsection{The model}

Consider three different harmonic oscillators,
\begin{equation}
H_i={1 \over 2} \left(p_i^2 + \omega_i^2 x_i^2\right)
\end{equation}
and define the two ``Hamiltonian constraints'', 
\begin{eqnarray}
H^{(0)}(1) &=& H_1 +H_2-H_3=0\\
H^{(0)}(2) &=& H_1 -H_2+2 H_3=0.
\end{eqnarray}
The two constraints have vanishing Poisson brackets among
themselves. We will have to request that the perturbed Hamiltonians
also have vanishing Poisson brackets for all values of $\lambda$ and
this will impose restrictions on our perturbative approach. 

As we discussed in the previous section, the eigenstates of the
quantum version of the constraints are simply given in the number
representation by $|n_1,n_2,n_3>$,
\begin{eqnarray}
H^{(0)}(1) |n_1,n_2,n_3>&=&
\hbar\left(\omega_1 n_1+K_1 +\omega_2 n_2+K_2
-\omega_3 n_3-K_3\right)|n_1,n_2,n_3>\\  
H^{(0)}(2) |n_1,n_2,n_3>&=&
\hbar\left(\omega_1 n_1+K_1 -\omega_2 n_2-K_2
+2 \omega_3 n_3+2 K_3\right)|n_1,n_2,n_3>.
\end{eqnarray}
It is possible to choose the $K_i$ in such a way that $|0,0,0>$ has
zero eigenvalue. The system admits other solutions only if 
\begin{equation}
\omega_2 = {M\over N} \omega_1,\quad \omega_3={P\over Q} \omega_1
\end{equation}
which leads to a system of equations for the $n_i$,
\begin{eqnarray}
n_1 +{M\over N} n_2 -{P\over Q}n_3&=&0\\
n_1 -{M\over N} n_2 +2 {P\over Q}n_3&=&0,
\end{eqnarray}
that has as general solution,
\begin{equation}
n_1=\ell P M,\quad n_2=-3 \ell P N,\quad n_3=-2\ell QM,
\end{equation}
with $\ell$ an arbitrary integer. Therefore the physical states of the
theory (states annihilated by the constraints) are,
\begin{equation}
|\ell P M, -3\ell N P, -2 \ell Q M>,\qquad {\rm integer}\, \ell.
\end{equation}

We will now consider a perturbative term that still allows the system
to be solved exactly,
\begin{equation}
H'_3=H_3+{\lambda \over 2}\omega_3^2 x_3^2,
\end{equation}
so, 
\begin{eqnarray}
H(1)&=&H^0(1)-{\lambda \over 2}\omega_3^2 x_3^2,\\
H(2)&=&H^0(2)+{\lambda \over 2}\omega_3^2 x_3^2.
\end{eqnarray}

The modification introduced by the perturbation is equivalent to
changing $\omega_3$ by $\omega_3'=\sqrt{1+\lambda}\omega_3$. If we now
consider the construction of quantum states $|n_1,n_2,n_3>$ that are
annihilated by both Hamiltonian constraints, and seek for values of
$n_i$ that differ only slightly from the unperturbed ones, which
is achieved by choosing $p,q,s\ll \ell$,
\begin{equation}
n_1=\ell PM+p,\quad n_2=-3NP \ell+q,\quad n_3=-2\ell Q M +s
\end{equation}
we find that, 
\begin{eqnarray}
p+{M\over N}q-(\sqrt{1+\lambda}-1){P\over Q}(-2\ell M
Q)-\sqrt{1+\lambda}{P\over Q} s&=&0\\
p-{M\over N}q+2(\sqrt{1+\lambda}-1){P\over Q}(-2\ell M Q)+2
\sqrt{1+\lambda} {P\over Q}s&=&0.
\end{eqnarray}
The joint solution of these equations requires that 
$p=M b, q=-3Nb$, and assuming that $\lambda$ is small, we find that,
\begin{equation}
2 M b-{P\over Q} s + {\lambda \over 2} (2 \ell M P-{P\over Q} s)=0
\end{equation}
which implies that 
\begin{equation}
\lambda={s \over \ell M Q}-{2 b\over \ell P} + O({1 \over \ell^2}).
\end{equation}

\subsection{Perturbative treatment}

We recall the form of the unperturbed Hamiltonian and the first order
perturbations,
\begin{eqnarray}
H^{(0)}(1)&=&H_1+H_2-H_3,\qquad H^{(1)}=-{1\over 2}\omega_3^2 x_3^2,\\
H^{(0)}(2)&=&H_1+H_2+2H_3,\qquad H^{(1)}=2{1\over 2}\omega_3^2 x_3^2,\\
\end{eqnarray}
with $\omega_2={M\over N} \omega_1$ and $\omega_3={P\over Q}
\omega_1$, and we wish to solve for quantum states,
\begin{equation}
<\phi^{(0)}|+\lambda<\phi^{(1)}| \left(H^{(0)}(a)+\lambda
H^{(1)}(a)\right)=
\left(\epsilon^{(0)}_a+\lambda \epsilon^{(1)}_a\right) 
\left(<\phi^{(0)}|+\lambda<\phi^{(1)}|\right),\qquad a=1,2.
\end{equation}
The spectrum of $H^{(0)}$ is given by,
\begin{eqnarray}
\epsilon^{(0)}_1 &=& \hbar \omega_1 \left[ n_1 +{M\over N} n_2 -{P\over 
Q} n_3\right]+\hbar\left(K_1+K_2-K_3\right),\\
\epsilon^{(0)}_2 &=& \hbar \omega_1 \left[ n_1 -{M\over N} n_2 +2{P\over 
Q} n_3\right]+\hbar\left(K_1-K_2+2K_3\right).
\end{eqnarray}
Once more, we have a degenerate spectrum, with,
\begin{equation}
n_1=\ell P M +p,\qquad n_2 =-3\ell N P+q,\qquad n_3=-2\ell QM +s,
\end{equation}
so the energies are,
\begin{eqnarray}
\epsilon^{(0)}_1(p,q,s)&=&
\hbar \omega_1 \left[p+{M\over N}q-{P\over Q}s\right]+
\hbar\left(K_1+K_2-K_3\right)\\ 
\epsilon^{(0)}_2(p,q,s)&=&
\hbar \omega_1 \left[p-{M\over N}q+2{P\over Q}s\right]+
\hbar\left(K_1-K_2+2K_3\right),
\end{eqnarray}
and the eigenstates are,
\begin{equation}
<\psi^{(0)}_{\ell,p,q,s}|=<\ell P M +p,-3 \ell N P+q,-2\ell Q M + s|.
\end{equation}

Going to next order in perturbation theory, we again have to require
(as in the example we discussed in the previous section) that the
first order correction to the state satisfy,
$<\phi^{(0)}|\phi^{(1)}>=0$. We now compute the first order correction
to the eigenvalues,
\begin{equation}
<\phi^{(0)}| H^{(1)}(a)|p,q,s,\ell>=\epsilon^{(1)}_a <\phi^{(0)}|p,q,s,\ell>,\quad a=1,2,
\end{equation}
or, explicitly,
\begin{eqnarray}
\epsilon^{(1)}_1 &=& -{1 \over 2} \hbar \omega_3
<\phi^{(0)}|\hat{\bar{x}}_3^2 |p,q,s,\ell> =-{1\over 2} \hbar \omega_3
\left(-2\ell QM+s +{1 \over 2}\right),\\
\epsilon^{(1)}_2 &=& \hbar \omega_3
<\phi^{(0)}|\hat{\bar{x}}_3^2 |p,q,s,\ell> = \hbar \omega_3
\left(-2\ell QM+s +{1 \over 2}\right).
\end{eqnarray}

As before, we choose the $K_i$'s in such a way that $|0,0,0>$ is a solution,
\begin{eqnarray}
K_1 +K_2-K_3-{\lambda \over 4} \omega_3&=&0\\
K_1 -K_2+2K_3+{\lambda \over 2} \omega_3&=&0.
\end{eqnarray}

We now require that the states have vanishing eigenvalues,
\begin{eqnarray}
\epsilon^{(0)}_1+\lambda \epsilon^{(1)}_1 &=& 
\hbar \omega_1 \left(p+{M\over
N} q -{P\over Q} s-{\lambda \over 2} {P\over Q} (-2 \ell Q
M+s)\right)=0\\
\epsilon^{(0)}_2+\lambda \epsilon^{(1)}_2 &=& 
\hbar \omega_1 \left(p-{M\over
N} q +2{P\over Q} s+{\lambda } {P\over Q} (-2 \ell Q M+s)\right)=0.
\end{eqnarray}

For these equations to have a solution with a common value of
$\lambda$, we need to choose carefully the states $<p,q,s,\ell|$, i.e,
we must have $q=-3N b, p=M b$, and the solution for $\lambda$ is,
\begin{equation}
\lambda\sim -{2 b\over P\ell} + {s\over QM\ell} \label{forla}
\end{equation}
so we see we completely reproduce the expansion of the exact solution
we found in the previous subsection. As in the previous example,
there are many eigenstates one can start from in the zeroth order
theory that yield a solution for the same $\lambda$. The formula
(\ref{forla}) is invariant if one multiplies $b,s,l$ times an
arbitrary integer.

We therefore see that indeed one needs extra conditions in order to
have both constraints vanish on a given value of the perturbative
parameter. But the conditions just imply that one starts with
different zeroth order eigenvalues for the states in both constraints.
The situation is schematized in the figure.

\begin{figure}
\centerline{\psfig{figure=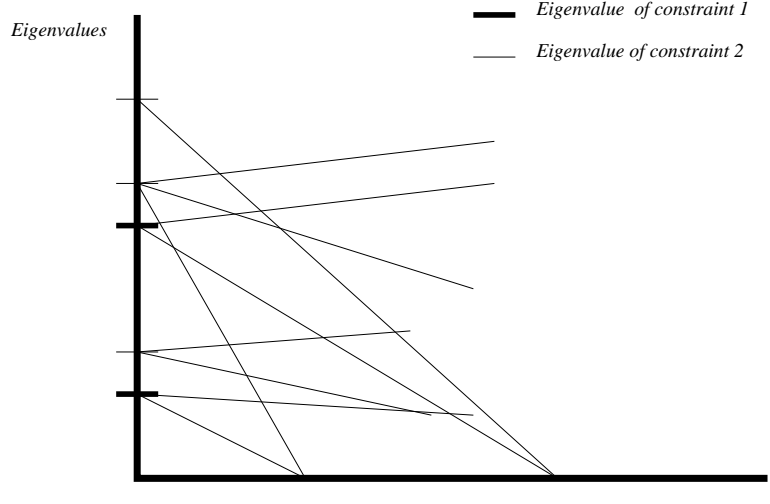,height=70mm}}
\caption{A schematic diagram of how the constraints are solved. One
starts from the eigenvalues of the unperturbed constraints of the
system. The perturbation breaks the degeneracy of the eigenvalues and
changes linearly the eigenvalues. When the corrected eigenvalues cross
the zero value simultaneously for both constraints, one has a
permitted value of $\lambda$.}
\end{figure}

Finally, if one sought to compute the first order correction of the
quantum state, one get two equations for it, one per each
constraint. If one studies them in detail, one finds that they admit
the same solution.

\section{Sketching the full quantum theory}

We do not have a detailed discussion of the full general relativity
quantum theory ready, but nevertheless one can sketch how the
procedure would be. The calculations in this section should only be
taken as a guide, and will be highly formal in nature. The details are
very important in determining the final quantum theory, therefore we
can only refer to these calculations as a ``sketch'' of what can be
achieved. We will see simultaneous solutions to the diffeomorphism and
Hamiltonian constrain. In order to achieve this we start with a basis
consistent of equivalence classes of spin nets under subgroup of
diffeomorphisms that keeps  the vertices of the spin net fixed,
\begin{equation}
|s_{[\vec{v}]}> =\sum D_{[\vec{v}]} | s>,
\end{equation}
where the $D_{[\vec{v}]}$ are the diffeomorphisms that leave the
vertices $\vec{v}$ of the spin network fixed. This is a space very
similar to the ``habitat'' considered by Lewandowski and Marolf
\cite{LeMa}. The action of both the diffeomorphism constraint and
Thiemann's Hamiltonian is well defined in such a space. We can
therefore proceed to study the problem of eigenvalues and eigenvectors
to first order in perturbation theory,
\begin{equation}
\left(<\psi^{(0)}|+\Lambda^{-1} <\psi^{(1)}|\right) \left(\hat{V}(M)+
\Lambda^{-1} \hat{H}(M)\right) = 
\left(<\psi^{(0)}|+\Lambda^{-1} <\psi^{(1)}|\right)\left(\epsilon^{(0)}(M)+
\Lambda^{-1} \epsilon^{(1)}(M)\right),
\end{equation}
Considering the equation at zeroth order we get,
\begin{equation}
<\psi^{(0)}| V(M)  = \epsilon^{(0)}(M)<\psi^{(0)}|
\end{equation}
the solution of this equation is given by the eigenvalues and
eigenvectors of the determinant of the metric. Here we have to deal
with the fact that this is a field theory. One can smear the
determinant of the metric with a function $M(x)$. The resulting
operator is closely related to the well-understood
\cite{RoSmvo,AsLevo} volume operator, we call it ``smeared volume''
and denote it as $V(M)$. One will essentially get the volume
associated with each vertex, which is a finite quantity, multiplied
times the value of the smearing function at the vertex. Eigenstates
will be reasonably easy to find. They will correspond to combinations
of spin networks that yield the same total for the sum of the volumes
of each vertex times the smearing function valued at the vertex.  That
quantity, in turn will be the eigenvalue $\epsilon^{(0)}(M)=\sum_{v_i}
V(v_i) M(v_i)$.  We can schematically write these states as
$<\psi^{(0)}_{v}| = \sum_{s} C(s)_v <\{s\}_{v,\epsilon^0(M))}|$ where we
have a sum of states such that the all have the same value of $V(M)$
for a given $M$.  There are many states for each value of
$\epsilon^{(0)}(M)$, to illustrate this, consider a spin network with
a given number of vertices and a given value of $\epsilon^{(0)}(M)$
and replace two lines in it by the same two lines with some knotting
in between them. The value of the eigenvalue will not change, but it
will be a different spin network.

Let us now consider the first order corrections. As we did in the
quantum mechanical examples, we will project the perturbative equation
on the space of states $<s_{V(M)}|$, 
\begin{equation}
<\psi^{(1)}|\hat{V}(M)|s_{V(M)}> + 
<\psi^{(0)}_{\epsilon^{(0)}(M)}|
\hat{H}(M)| s_{V(M)}> =\epsilon^{(1)}(M)
<\psi^{(0)}_{\epsilon^{(0)}(M)}|s_{V(M)}>+
\epsilon^{(0)}(M)
<\psi^{(1)}|s_{V(M)}>.
\end{equation}
The first and last term in both members cancels out, since
$\hat{V}(M)|s_{V(M)}>=
\epsilon^{(0)}(M)|s_{V(M)}>$. One is therefore left with,
\begin{equation}
<\psi^{(0)}_{\epsilon^{(0)}(M)}|
\hat{H}(M)| s_{V(M)}> =\epsilon^{(1)}(M)
<\psi^{(0)}|s_{V(M)}>.
\end{equation}
One can use this equation to explicitly compute $\epsilon^{(1)}(M)$,
for a given proposal for the Hamiltonian constraint. One can now
proceed to make the ``energy'' vanish at this order of perturbation
theory, therefore determining the (quantized) 
value of the cosmological constant and the possible values of the 
energy of the zeroth order state $\epsilon^0(M)$.

To obtain the correction to the state, we need to project the above
equation on states that are orthogonal to the eigenstates with
eigenvalue $\epsilon^{(0)}(M)$. Let us call
those states $|s_{V'(M)}>$, that is, states with different values of
the volume at the vertices.   We therefore  assign the volume at the
vertices as a measure of orthogonality. We therefore write,
\begin{equation}
<\psi^{(1)}|\hat{V}(M)|s_{V'(M)}> + \Lambda^{-1}
<\psi^{(0)}_{\epsilon^{(0)}(M)}|
\hat{H}(M)| s_{V'(M)}> =\epsilon^{(0)}(M)
<\psi^{(0)}_{\epsilon^{(0)}(M)}|s_{V'(M)}>+\Lambda^{-1}
\epsilon^{(0)}(M)
<\psi^{(1)}|s_{V'(M)}>,
\end{equation}

Now, since $\hat{V}(M)|s_{V'(M)}>={\epsilon'}^{(0)}(M)s_{V'(M)}>$,
we can write,
\begin{equation}
<\psi^{(1)}|s_{V'(M)}> = {<\psi^{(0)}_{\epsilon^{(0)}(M)}|
\hat{H}(M)| s_{V'(M)}> \over {\epsilon'}^{(0)}(M)-\epsilon^{(0)}(M)}.
\end{equation}
and this equation would determine the first order correction to the
quantum state. This would be only strictly true if the state of volume
$V(M)$ were non-degenerate. The degeneracy implies that there might be
non-vanishing projections $<\psi^{(1)}|s_{V(M)}>$ for  vectors
that belong to the space of states with eigenvalue
$\epsilon^{(0)}(M)$. It is well known in degenerate perturbation
theory that higher order calculations determine these components.

The dependence on $M$ of the last equation might appear as surprising,
since the first order quantum state should be $M$ independent. Here we
can draw on our experience on the system with two constraints studied
in last section, which shows that a single correction appears no
matter which Hamiltonian one uses to compute the correction. In the
gravity case this would correspond to getting the same solution for
different $M$'s. Another way to see that the equation is
$M$-independent is to notice that one can characterize the action of
the Hamiltonian independently at each vertex and in such calculations
the role of $M$ is that of a constant overall factor that  drops
out of the equations and one is left, as in the case of the
oscillators, with a system of equations.

We have therefore constructed states such that,
\begin{equation}
<\psi_v|(\Lambda) = <\psi^{(0)}_v|+\Lambda^{-1} <\psi^{(1)}_v|
\end{equation}
such that,
\begin{equation}
<\psi_v(\Lambda)| H(M,\Lambda) = O(\Lambda^{-2}) \forall M.
\end{equation}

One is however, interested in states that are genuinely invariant
under diffeomorphisms, whereas the above states are only invariant
under diffeomorphisms that leave the vertices of the spin network
fixed.  The kind of states we look for are,
\begin{equation}
<\psi(\Lambda)| =\sum_{\cal D} <\psi_v (\Lambda)| {\cal D},
\end{equation}
such that 
\begin{equation}
<\psi(\Lambda)| H(M,\Lambda) = O(\Lambda^{-2}).
\end{equation}
We shall show that with the above defined states this is indeed the
case. This follows from the Poisson algebra of diffeomorphism and
Hamiltonian constraints,
\begin{equation}
{\cal D} H(M,\Lambda) {\cal D}^{-1} = H({\cal D} M,\Lambda).
\end{equation}
Therefore,
\begin{equation}
<\psi(\Lambda)| H(M,\Lambda) =\sum_{\cal D} <\psi_v (\Lambda)|
{\cal D} H(M,\Lambda) = \sum_{\cal D} <\psi_v (\Lambda)|
H({\cal D} M,\Lambda) {\cal D}=O(\Lambda^{-2}),
\end{equation}
which is what we wished to show.

The details of these calculations can only be filled in with a given
prescription for a Hamiltonian constraint and a definite space of
states to operate upon. This will require much more detailed work than
is appropriate for this paper. One can hint at what would happen, for
instance, if one uses Thiemann's Hamiltonian. Let us start with
trivalent spin networks. In such a context $V(M)$ vanishes identically
and the perturbative method has nothing to add: one is left without
higher order corrections and the solution to the problem is reduced to
determining which states are annihilated by the Hamiltonian
constraints. To get something non-trivial, one needs four-valent
vertices.  Calculations on four valent vertices are readily possible,
but cumbersome. In this context one could sharpen the determination of
eigenstates, etc. Very schematically, one would see that the
correction to the states that one gets depend on the action of the
Hamiltonian constraint at each vertex. The states appear different
from the ones obtained via ``group averaging'' that involve acting
repeatedly with the Hamiltonian. However, since the action of the
Hamiltonian is ``local'' (i.e. it does not add connections among
vertices, but ``dresses up'' each vertex), the kind of states one gets
have similar locality properties to those obtained with the ``group
averaging'' procedure. In fact, if one works at higher order, at each
order one acts with a Hamiltonian, so one can see that by going to
higher orders, one is recovering something similar to the ``group
averaging''. 

An interesting point in trying to draw an analogy between quantum
gravity and the simple models discussed in this paper is if the
perturbation Hamiltonian has a discrete spectrum or not. We know that
the zeroth order Hamiltonian (the smeared volume) has a discrete
spectrum. At the moment, it is knot known if either the Thiemann or
the Vassiliev Hamiltonians have a discrete spectrum. However, we have
shown in this paper that the perturbative method works well in the
case in which the perturbative Hamiltonian has a continuous spectrum
(the case of the coupled oscillators). It is well known that the
perturbative approach works well when the perturbing Hamiltonian has a
discrete spectrum (for instance if one considers an atom in a magnetic
field with constant energy).

It is clear that this is just the beginning of a discussion of the
four dimensional quantum gravity case. It is interesting to notice,
however, how one can quickly gain intuition as to the kind of states
one would get and how they depend on the action of the Hamiltonian
without doing the explicit computations. Since the latter must involve
four valent vertices and therefore are extensive in nature, having a
quick way to intuitively try out proposals for the Hamiltonian is a
great asset.

\section{Conclusions}

It is obvious that the complexities that one expects in quantum
gravity (in particular the infinite dimensional nature of the problem)
cannot be really mimicked by finite dimensional systems. We believe,
however, that the finite dimensional models we analyze in this paper
help dissipate some of the most elementary skepticisms about the
possibilities of the approach we are proposing. Namely: 1) that the
method seems to simplistic to deal with chaotic, possibly pathological
Hamiltonian systems as one expects full general relativity to contain
in certain regimes; 2) that the method is only circumscribed to a
range of values of the cosmological constant of no physical relevance;
3) that the application in the quantum domain of the technique is
possibly inconsistent. 

Applied to the full quantum theory, our method appears to reduce the
problem of finding quantum states to a set of well defined, albeit
complicated spin network calculations. Moreover, it appears to yield a
quick intuitive handle on the properties of the quantum states of
possible proposals for Hamiltonian constraints.  We are currently
exploring further the application of the approximation technique in
full quantum gravity, as sketched in the last section. 

\acknowledgements This paper was inspired by Karel Kucha\v{r}'s
suggestions that we test our proposal in several model examples, some
of which he pointed out; he also made valuable criticisms to the
manuscript. We are grateful to Abhay Ashtekar and Charles Torre for
discussions. This work was supported in part by the National Science
Foundation under grant INT-9811610, and research funds of the
Pennsylvania State University.  We acknowledge support of PEDECIBA. We
wish to thank Thomas Thiemann and the Albert Einstein Institute of the
Max Planck Society for hospitality during the completion of this work.

\end{document}